\documentclass[preprint,12pt]{elsarticle}

\usepackage{amssymb}

\journal{Physics Letters A}

\begin{document}

\begin{frontmatter}


\ead{vergara@nucleares.unam.mx}

\title{Hidden Gauge Symmetry in Holomorphic Models}


\author{Carlos A. Margalli and J. David Vergara}

\address{Instituto de Ciencias Nucleares,\\
\it Universidad Nacional Aut\'onoma de M\'exico,\\
\it A. Postal 70-543 , M\'exico D.F., M\'exico}

\begin{abstract}
We study the effect of a hidden gauge symmetry on complex holomorphic systems. For this purpose, we show that intrinsically any holomorphic system has this gauge symmetry. We establish that this symmetry is related to the Cauchy-Riemann equations, in the sense that the associated constraint is a first class constraint only in the case that the potential be holomorphic. 
As a consequence of this gauge symmetry on the complex space, we can fix the gauge condition in several ways and project from the complex phase-space  to real phase space. Different projections are gauge related on the complex phase-space but are not directly related on the real physical phase-space.        

\end{abstract}


\end{frontmatter}



\section{Introduction}
In several instances, in physics it is natural to select complex variables to develop a theory. For example, in Conformal Field Theory in two dimensions the conformal transformations of the metric are equivalent to the Cauchy-Riemann equations for holomorphic functions.  In this paper, we consider a generalization of this concept, in the sense that, we regard a system defined in the complex space and we show that this system possesses a gauge symmetry. This symmetry is trivial when is analyzed directly in the context of the complex variables $z=x+iy$, because it says directly that the transformation is null $\delta z=0$, then all the holomorphic functions are invariant under these transformations. However, these transformations are not trivial if we consider that the real and imaginary part of $z$ are allowed to transform on the complex plane. We show that these transformations are gauge transformations generated by a first class constraint in the context of the Dirac's canonical method. Then by selecting a gauge condition we can map our complex system to different real systems. The interesting point is that these systems are related by a gauge transformation on the complex phase-space.

In Quantum Mechanics one of the fundamental postulates is that every measurable physical quantity ${\mathcal A}$ is described by an operator $A$ acting on the state space; and this operator is an observable. A common hypothesis is to select Hermitian operators, in order to obtain measurable or observable quantities. This postulate implies that if we want to get all the information of the system, we must consider a complete set of commuting observables. Moreover,  it has been hypothesized that some systems do not necessarily satisfy this postulate. Examples of these cases are: non-Hermitian models \cite{Moise, Most.1}, with interesting applications as to generate entanglement in many-body systems \cite{Moise.1}. The PT-symmetry \cite{Bender.1, Most}, with striking applications in optics \cite{Muss,Vyslo}. Also, we have theories with high order time derivatives as the Pais-Uhlenbeck model for particles  \cite{Pai, Smil, Most.2}   and Bernard- Duncan model for fields  \cite{Bernard}, noncommutative theories  \cite{witten, Alvarez}, higher order derivative theories of gravity  \cite{stelle, stelle.2} and complex theories of gravity \cite{Ashtekar.b}.

There are several ways to address these models, for example when the Hermiticity is not available, it is natural to introduce a new kind of symmetry and in this way, the notion of PT-symmetry was introduced by Bender \cite{Bender.1}.  Furthermore, Ashtekar introduced a modification of the internal product, using the reality conditions, and this procedure also solves the problem in some cases \cite{Ashtekar.b}. Our approach is a generalization of  the Ashtekar procedure, but written in a different way. Some years ago was shown that the reality conditions can be interpreted as second class constraints in the context of the Dirac's method of canonical quantization \cite{Morales}, and then the internal product is given in terms of the measure of the path integral with second class constraints. The object of this paper is to explore further this idea. We find that in any holomorphic theory there is intrinsically a gauge symmetry and the second class constraints of the Ashtekar formalism correspond to a selection of the gauge condition of the symmetry. However, there are many additional consistent gauge conditions. By selecting a gauge condition we get a different real physical system that is gauge related to another real system by a complex gauge transformation to be performed on the extended complex phase-space. In this form, we will show that using this gauge symmetry we can relate on the complex phase-space different real systems that are not related by a canonical transformation on the real phase-space. The work is organized as follows. In Sec. 2, we introduce the gauge symmetry and we show that is related to the Cauchy-Riemann equations. In Sec. 3 starting from the complex harmonic oscillator, and by using different gauges we obtain in the real physical space the potentials $k/2 x^2,\ a x^{-1},\ b x^{-2}$ and $-a x^{-1}$. In Sec. 4 we generalize our construction to the two-dimensional complex space and we show that the harmonic oscillator and the Kepler problem are gauge related. In Sec. 5 we quantize  the system using path integrals. Section 6 is devoted to our conclusions.

\section{Complex Theory for a First Order Theory}\label{esco}
Let us consider a complex Lagrangian that is a function of the holomorphic coordinate  $z=x+iy$ and their velocities 
\begin{eqnarray}\label{lagrangian1}
 L(z,\dot z)=\frac{1}{2}\dot{z}^{2}-V(z)\end{eqnarray}
and we are assuming that the potential $V(z)$ is a holomorphic function of $z$, that is,
\begin{equation}\label{CRE6}
\frac{dV}{d\bar z}=0
\end{equation}
Then, it is evident that the Lagrangian is  invariant under the transformations
\begin{eqnarray}\label{finitet}
 x'=x+\lambda(t),\qquad y'=y+i\lambda(t).
\end{eqnarray}
That leave $z$ invariant, i.e. $\delta z=0$. In consequence, from this point of view our system have a trivial symmetry.  On the other hand, this symmetry is more useful if  we decompose $z$ in terms of real and imaginary parts. In this case, the Lagrangian is given by 
\begin{equation}
L(z,\dot z)=\frac{1}{2}\dot{x}^{2}-\frac{1}{2}\dot{y}^{2}+i\dot{x}\dot{y}
-V_{R}(x,y)-iV_{I}(x,y).
\end{equation}
The associated equations of motion of the above Lagrangian are not independent, since it is possible to divide them in real and
imaginary parts and we get
\begin{equation}
 \ddot{x}+\frac{\partial V_{R}(x,y)}{\partial x}=0,\label{eqmov}\qquad
\ddot{y}-\frac{\partial V_{R}(x,y)}{\partial y}=0,
\end{equation}
where it is clear that $x$ and $y$ are real quantities. Now, we proceed to develop the canonical formulation of this theory using the variables $x$ and $y$. We select these variables,  because in terms of holomorphic coordinates $z$, our symmetry is trivial in the sense we can not establish, any relation between the holomorphic and anti-holomorphic coordinates.
The canonical momenta for the Lagrangian (\ref{lagrangian1}), are
\begin{eqnarray}
 p_{x}:=\frac{\partial L}{\partial \dot{x}}=\dot{x}+i\dot{y},\qquad \qquad p_{y}:=\frac{\partial L}{\partial \dot{y}}=-\dot{y}+i\dot{x}.\end{eqnarray}
Using these definitions, we obtain the primary constraint
\begin{equation}\label{primarycons}
\Phi = 2 p_{\bar z}= p_{x}+ip_{y}\approx0,
\end{equation}
where we introduce the weak equality symbol "$\approx$" to emphasize that the quantity $\Phi$ is numerically restricted to be zero but does not identically vanish throughout phase space \cite{Hen}.
Following the usual definition of the canonical Hamiltonian in the phase-space
\begin{eqnarray}\label{constraint}
 H=\dot{x}p_{x}+\dot{y}p_{y}-L,
\end{eqnarray}
we observe that $H, L, p_{x}, p_{y}$ are complex quantities.
Through the definition (\ref{constraint}) we obtain the explicit total Hamiltonian
\begin{equation}
 H_{T}=\frac{p_z^2}{2}+V(z)+\mu \Phi=\frac{1}{2}p_{x}^{2}+V_{R}(x,y)+iV_{I}(x,y)+\mu \Phi,
\end{equation}
where we add the primary constraint following the Dirac's method \cite{Dir}. The resulting equations of motion are
\begin{eqnarray}\label{pos0}
 \dot{x}=\{x,H_{T}\}=p_{x}+\mu,\qquad \dot{y}=\{y,H_{T}\}=p_{y}+i\mu,\\ \label{pos3}
\dot{p}_{x}=\{p_{x},H_{T}\}=-\frac{\partial V}{\partial x},\qquad
\dot{p}_{y}=\{p_{y},H_{T}\}=-\frac{\partial V}{\partial y}. 
\end{eqnarray}
Using equations (\ref{pos0}) we observe that there is a gauge freedom since we have an
 arbitrary Lagrange multiplier. It is important to note that the temporal evolution is not necessarily a real quantity,
so our real variables $x$ and $y$ under evolution could obtain an imaginary contribution.

In the following, we are going to evolve the primary constraint (\ref{primarycons}) and in this way we should understand what kind of constraint is,
first or second class. The primary constraint $\Phi$ evolves as
\begin{eqnarray}
 \dot{\Phi}=\{\Phi,H\}=-2\frac{dV(z)}{d{\bar z}}\approx 0.
\end{eqnarray}
In the present case, we observe that the temporal evolution of
$\Phi$ imposes as a result the  Cauchy-Riemann equations for $V(x,y)$ (see Eq. (\ref{CRE6})). In consequence if $V(x,y)$ holomorphic function we obtain that $\Phi$ is first class constraint.
Furthermore, there are no additional constraints and we will get as a result that our reduced phase-space has therefore two degrees of freedom. In the other hand, following the Dirac's quantization method the physical states are defined by imposing that the action of the first class constraint over the states is equal to zero. In the coordinate representation this we will imply   
\begin{eqnarray}
\left(-i\hbar\frac{\partial}{\partial x}+\hbar\frac{\partial}{\partial y} \right)[G_{R}(x,y)+iG_{I}(x,y)]=0,
\end{eqnarray}
resulting the  Cauchy-Riemann equations, if we decompose in real and imaginary parts.
In this way, the Cauchy-Riemann equations appear in this formalism
as an invariance under translations generated by the constraint.
In other words, we obtain that our theory is compatible with
the Dirac's formulation \cite{Dir,Hen}, but it must satisfy that the potential is a holomorphic function
\begin{eqnarray}
 \hat{\Phi} V(x,y)=0.
\end{eqnarray}

Now, following the Dirac's conjecture this constraint will be the generator of gauge transformation \cite{Dir}. 
The transformations produced by the first class constraint are
\begin{eqnarray}\label{cfc}
\delta x=\{x,\epsilon \Phi\}=\epsilon,\qquad
\delta y=\{y,\epsilon \Phi\}=i\epsilon,\label{multiplier}\\
\delta p_{x}=\{p_{x},\epsilon \Phi\}=0,\qquad
\delta p_{y}=\{p_{y},\epsilon \Phi\}=0.\nonumber\\
\end{eqnarray}
In consequence, we get
\begin{equation}
 \delta z=\{z, \Phi\}=0,
\end{equation}
in agreement with the transformations (\ref{finitet}). From a pragmatic point of view $z$  and $p_{z}$ are Dirac's observables with null variation, but it implies a more complicated structure
if we take in account the variation of the real and imaginary part of $z$.  In this framework and 
if we pay attention to the equations (\ref{pos0}),
it is necessary to impose a gauge condition in order to obtain a real reduced phase-space.
Then according to the gauge condition that we choose we can obtain a different real theory. 
The interesting point is that all these real theories are connected by a complex gauge transformation in the original extended phase-space.
In fact, the equations of motion that we get from the Hamiltonian formulation are complex quantities obtained for the real and imaginary parts.
Furthermore, the phase-space is wider than the configuration space since
 $\mu$ exists in this  formulation, and it is possible to choose as a real quantity either
$\delta x$   or  $\delta y$. For the purpose of applying a method that is not trivial using this mathematical structure,
we must propose a gauge condition on the real or imaginary part of $z$ in such way that the resulting theory be real and in consequence the imaginary degrees of freedom be eliminated.
Furthermore, this will produce that the components, real and imaginary, interact with each other and send information from one side to the other. 
In addition, the procedure is not forcing us to fix the gauge symmetry for the Hamiltonian in a specific way and the gauge condition is selected
according to the  real model that we wish to build. It shares a common origin with the Ashtekar's complex models
based on the reality conditions and the imposition of the gauge condition is equivalent to the introduction of the internal product \cite{ashtekar}.
However, the reality conditions are a single way to select a gauge condition. Whereas, there are a lot of ways
to remove this freedom in the complex systems so we get to a Hermitian system.

The next step is to determine the Lagrange multiplier $\mu$ (\ref{pos0}), using the gauge condition $\gamma(x,y)$.  The canonical evolution of this condition gives us
\begin{eqnarray}
 \dot{\gamma}=\{\gamma,H_{T}\}=\{\gamma,H+\mu \Phi\}=\{\gamma,H\}+\mu\{\gamma,\Phi\}\approx0,
\end{eqnarray}
then
\begin{eqnarray}
 \mu\approx -\frac{\{\gamma,H\}}{\{\gamma,\Phi\}}=-\frac{p_{x}\frac{\partial\gamma}{\partial x}}{(\frac{\partial\gamma}{\partial x}+i
\frac{\partial\gamma}{\partial y})}.
\end{eqnarray}
This implies for the equations of motion the following expressions
\begin{eqnarray}
 \dot{x}=[1-\frac{\frac{\partial\gamma}{\partial x}}{(\frac{\partial\gamma}{\partial x}+i
\frac{\partial\gamma}{\partial y})}]p_{x},\qquad\dot{y}=-i\frac{p_{x}
\frac{\partial\gamma}{\partial x}}{(\frac{\partial\gamma}{\partial x}+i
\frac{\partial\gamma}{\partial y})},\label{pos}\\
\dot{p}_{x}=\{p_{x},H_{T}\}=-\frac{\partial V_{R}(x,y)}{\partial x}
-i\frac{\partial V_{I}(x,y)}{\partial x},\nonumber\\
\dot{p}_{y}=\{p_{y},H_{T}\}=-\frac{\partial V_{R}(x,y)}{\partial y}
-i\frac{\partial V_{I}(x,y)}{\partial y}.\label{moma}
\end{eqnarray}
Now, assume that the gauge condition is of the form
\begin{equation}\label{gamma11}
 \gamma_{1}=y-ig_{1}(x)\approx0
\end{equation}
and it determines a concrete value for $\mu$
\begin{eqnarray}
 \mu_{1}\approx -\frac{\{(y-ig_{1}(x)),H\}}{\{(y-ig_{1}(x)),\Phi\}}=-\frac{p_{x}
\frac{\partial g_{1}(x)}{\partial x}}{(\frac{\partial g_{1}(x)}{\partial x}-1)},
\end{eqnarray}
with the resulting equations of motion for the reduced phase-space,
\begin{eqnarray}\label{power1}
 \dot{x}=[1-\frac{\frac{\partial g_{1}(x)}{\partial x}}{(\frac{\partial g_{1}(x)}{\partial x}-1
)}]p_{x},\label{poso}\\
\dot{p}_{x}=\{p_{x},H_{T}\}=-\frac{\partial V_{R}(x,ig_1)}{\partial x}
-i\frac{\partial V_{I}(x,ig_1)}{\partial x},\nonumber
\end{eqnarray}
Now, if the potential is a power series in $z$ then
\begin{equation}
V(z)=az^n=V_R(x,y)+iV_I(x,y)
\end{equation}
for $n$ even or odd, we get 
\begin{equation}
V_R(x,y)\sim x^m y^l  \qquad V_I(x,y)\sim x^i y^j
\end{equation}
with  $n$ even,  $(m,l)$ are even and $(i,j)$ are odd. In the case that $n$ is odd, we get that $(m,j)$ are odd and $(l,i)$ are even.  Then by Eq. (\ref{power1}) the evolution of $p_x$ is real and the gauge condition (\ref{gamma11}) projects correctly. 
So, $x$ and $p_{x}$ are real quantities and they don't leave the real space, with the evolution.
Furthermore, using this procedure the imaginary part in the potential contributes
to the real part through the gauge condition. 

However the method described above is meaningful, if the power series of
$V(z)$ is multiplied by a real constant but if we have an imaginary constant in front of $V(z)$ the method does not work.
In order to confront this case, we need a new gauge condition
\begin{equation}
 \gamma_{2}=x-ig_{2}(y)\approx0.
\end{equation}
Now, the real quantities  will be $y$, $p_{y}$, $\dot{p}_{y}$ and $L,\ H$.
We obtain a similar situation to (\ref{pos}) and (\ref{moma})
as in the case of $\gamma_{1}$.  Also we can select another kind of gauge conditions as
\begin{equation}
y = ix + i{g_1}(x){\rm{ }} + i{g_2}(x)p_x + i{g_2}(x)p_x^2
\end{equation}
In this case, we must look at real solutions to the momentum $p_x$ in terms of velocities. For this specific example, the second order action gets a term with a velocity dependent metric.
Then by selecting an appropriated gauge condition we will get a map from the complex phase-space to the real physical space. 
Now, if we want to quantize the real theory  we can follow two paths, first we can compute the corresponding Dirac's brackets for the variables on the reduced phase-space and promote these brackets to commutators. The second procedure is to build the measure of the path integral using the Senjanovic procedure \cite{Senjanovic}.  In the next sections, we will describe by several examples how the procedure  works specifically.

\section{Complex Harmonic Oscillator and gauge related systems}\label{oscilad}
In order to describe how the procedure introduced works explicitly, we select the harmonic oscillator.
First, we consider a complex extension of the ordinary Lagrangian, given by
\begin{equation}\label{CHO.1}
 L=\frac{1}{2}\dot{z}^{2}-\frac{\omega^{2}}{2}z^{2}
\end{equation}
where $z$ is a complex variable that is separated into real and imaginary parts
\begin{equation}
 z=x+iy.
\end{equation}
The Lagrangian in terms of these variables is
\begin{eqnarray}\label{Lagrangianacom.11}
 L_{(x,y)}=\frac{1}{2}\dot{x}^{2}-\frac{1}{2}\dot{y}^{2}+i\dot{x}\dot{y}-\frac{\omega^{2}}{2}x^{2}+\frac{\omega^{2}}{2}y^{2}
-i\omega^{2}xy
\end{eqnarray}
and we obtain the momenta
\begin{eqnarray}
 p_{x}=\dot{x}+i\dot{y},\\
p_{y}=-\dot{y}+i\dot{x}.
\end{eqnarray}
We observe that these momenta are not independent, then generate the primary constraint
\begin{equation}\label{cpc}
 \Phi_{0}=p_{x}+ip_{y}\approx 0.
\end{equation}
The next step is to obtain the canonical Hamiltonian
\begin{equation}
 H_{0}=\dot{x}p_{x}+\dot{y}p_{y}-L_{(x,y)},
\end{equation}
and the total Hamiltonian will be
\begin{equation}\label{oaham.1}
 H_{T}=\frac{1}{2}p_{x}^{2}+\frac{\omega^{2}}{2}x^{2}-\frac{\omega^{2}}{2}y^{2}+i\omega^{2}xy+\mu^{0} \Phi_{0}.
\end{equation}
On the other hand, the Poisson brackets are
\begin{eqnarray}\label{acomplex}
\{x,p_{x}\}=1,\qquad \{y,p_{y}\}=1.
\end{eqnarray}
Now, the evolution of the constraint (\ref{cpc}) give us 
\begin{eqnarray}\label{acomplex.11}
\{\Phi_{0},H_{T}\}=0,
\end{eqnarray}
resulting that $\Phi_{0}$ is a complex first class constraint.

The gauge transformations induced by this constraint on the phase-space are
\begin{eqnarray}\label{innor}
\delta x=\{x,\mu \Phi_{0}\}=\mu,\qquad
\delta y=\{y,\mu \Phi_{0}\}=i\mu,\label{delta}\\
\delta p_{x}=\{p_{x},\epsilon^{0} \Phi_{0}\}=0,\qquad
\delta p_{y}=\{p_{y},\epsilon^{0} \Phi_{0}\}=0.\nonumber\\
\end{eqnarray}
Here we observe that for $\mu$ real, the gauge transformation over the real variable $y$ induces an imaginary part, then this part can influence the real part of the Lagrangian and in this form to modify the dynamics of the real system.  
As a simple example, we choose the gauge condition that leads to the usual Hermitian structure
\begin{equation}\label{gauge.46}
 \gamma_{0}=y\approx0.
\end{equation}
The Lagrange multiplier is obtained through Dirac's method and we get
\begin{eqnarray}
 \dot{\gamma}_{0}=\{\gamma_{0},H_{T}\}=\mu^{0}\{\gamma_{0}, \Phi_0\}=i\mu^{0}\approx0,
\end{eqnarray}
The equations of motion from (\ref{pos0}) and (\ref{pos3}) yield
\begin{eqnarray}
 \dot{x}=\{x,H_{T}\}=p_{x}+\mu^{0},\qquad
\dot{p}_{x}=\{p_{x},H_{T}\}=-\omega^{2}x+i\omega^{2}\gamma_{0}.
\end{eqnarray}
The second equation says that the time evolution for $p_{x}$ is real.
However, $\gamma_{0}$ must be a good gauge condition and this implies that together with the constraint (\ref{cpc}), must be a set of second class constraints $\chi_a=(\Phi_0, \gamma_0)$.
The  matrix $C_{ab}=\{\chi_a,\chi_b\}$ for  the constraints is
\begin{equation}
\mathcal{C}_{ab}=\left(
\begin{array}{rccl}
0 & -i \\
i & 0 \\
\end{array}
\right).
\end{equation}
with determinant
\begin{equation}
\det \left( \mathcal{C}_{ab}\right) =-1.
\end{equation}
Then, the gauge condition $\gamma_{0}$ implies for the momentum
\begin{eqnarray}
 p_{x}=\dot{x}+i\dot{\gamma}_{0}\approx p_{xR},
\end{eqnarray}
that it is a real quantity. The  Hamiltonian (\ref{oaham.1}) is reduced to
\begin{equation}
 H_{HO}=\frac{1}{2}p_{xR}^{2}+\frac{\omega^{2}}{2}x^{2}.
\end{equation}
and the corresponding  Lagrangian is the usual one
\begin{eqnarray}\label{HO}
 L_{HO}=\frac{1}{2}\dot{x}^{2}-\frac{\omega^{2}}{2}x^{2}.
\end{eqnarray}
So in this way, we get the real harmonic oscillator. Indeed, by fixing the gauge condition (\ref{gauge.46}) we recover the real harmonic oscillator.
Another alternative way to fix the  gauge condition is
\begin{eqnarray}\label{gc11}
\gamma_{1}=y-ix+iU^{\frac{1}{2}}(x)\approx 0,\label{connorma}\\
\{\Phi,\gamma_{1}\}=-\frac{i}{2U^{\frac{1}{2}}}\partial_{x}U(x).
\end{eqnarray}
We use this particular form of the gauge condition to get an interesting form for the Hamiltonian in the reduced phase space ( see Eq. (\ref{reducedH}) ). Now, we get the set of 
second class constraints $\chi_{1a}=(\Phi_0,\gamma_1)$,  with the corresponding matrix of second class constraints given by
\begin{equation}
\mathcal{A}_{ab}=\left(
\begin{array}{rccl}
0 & -\frac{i}{2U^{\frac{1}{2}}(x)}\partial_{x}U(x) \\
\frac{i}{2U^{\frac{1}{2}}(x)}\partial_{x}U(x) & 0 \\
\end{array}
\right).
\end{equation}
and the inverse matrix yields
\begin{equation}
\mathcal{A}^{ab}=\left(
\begin{array}{rccl}
0 & -2\frac{i}{\partial_{x}U(x)}U^{\frac{1}{2}}(x) \\
2\frac{i}{\partial_{x}U(x)}U^{\frac{1}{2}}(x) & 0 \\
\end{array}
\right).
\end{equation}
The associated determinant is
\begin{eqnarray}\label{detA}
\det \mathcal{A}_{ab}=-\frac{[\partial_{x}U(x)]^{2}}{4U(x)}.
\end{eqnarray}
We observe here, that  the gauge (\ref{connorma})  must be accessible,  this means that given any set of canonical 
variables there must exist a gauge transformation that maps the given set onto a set that satisfies (\ref{connorma}). Furthermore, the condition (\ref{connorma}) must fix the gauge completely and that is the case only if the determinant (\ref{detA}) is different from zero. If the determinant vanishes in some point, the gauge condition is not defined globally and we have a Gribov obstruction \cite{Gribov,Espo}.  If this is not the case, the corresponding Dirac's bracket is
\begin{eqnarray}
 \{x,p_{x}\}^{*}=\frac{2U^{\frac{1}{2}}(x)}{\partial_{x}U(x)},
\end{eqnarray}
we note that this gauge condition generates a
non-canonical transformation.
The Hamiltonian in the reduced phase space is
\begin{eqnarray}\label{reducedH}
 H_{R}=\frac{1}{2}p_{x}^{2}+\frac{\omega^{2}}{2}U(x),
\end{eqnarray}
and we obtain that the action in the reduced space is
\begin{equation}\label{reduced1}
 S_{1R}=\int\!dt[\frac{\partial_{x}U(x)}{2U^{\frac{1}{2}}(x)}\dot{x}p_{x}-H_{R}],
\end{equation}
From the variation with respect to $p_x$ we can get the momentum in terms of the velocities 
\begin{eqnarray}
 p_{x}=[\frac{\partial_{x} U(x)}{2U^{\frac{1}{2}}(x)}]\dot{x},
\end{eqnarray}
that it is consistent with the expression (\ref{poso}).
In consequence the action in the configuration space is
\begin{eqnarray}\label{potV}
 S_{1R}=\int\!dt\left(\frac{\partial_{x}U}{2U^{\frac{1}{2}}}\right)^{2}\left[\frac{1}{2}\dot{x}^{2}-\omega^{2}
\frac{2U^{2}}{(\partial_{x}U)^{2}}\right].
\end{eqnarray}
For the Lagrange's multiplier we get
\begin{eqnarray}\label{mu1}
 \mu_{1}=p_{x}\left[1-\frac{2U^{\frac{1}{2}}(x)}{\partial_{x}U(x)}\right]=\left[\frac{\partial_{x}U(x)}{2U^{\frac{1}{2}}(x)}-1\right]\dot{x},
\end{eqnarray}
Note that if we assume that  the gauge condition (\ref{gc11}) is infinitesimal related to the gauge condition (\ref{gauge.46}),
the associated gauge transformation that related one system with the other is 
\begin{eqnarray}
 \tilde{x}=x-i\delta\gamma \{x,\Phi\}, \qquad \tilde{x}=U^{\frac{1}{2}}(x),
\end{eqnarray}
where
\begin{eqnarray}
 \delta\gamma=\gamma_{1}-\gamma_{0}
\end{eqnarray}
with the objective of replacing $\tilde{x}$ with $x$ in (\ref{HO})
to obtain (\ref{potV}).

To rewrite the action (\ref{reduced1}) in more conventional way we introduce
the repa\-rametrization
\begin{eqnarray}\label{repara.1}
 (\frac{d\tau}{dt})=[\frac{4U}{(\partial_{x}U)^{2}}],
\end{eqnarray}
then the action (\ref{reduced1}) is transformed to
\begin{eqnarray}\label{lagrangered}
 S=\int\!d\tau [\frac{1}{2}(x')^{2}
-\frac{\omega^{2}}{8}(\partial_{x}U)^{2}].
\end{eqnarray}
for the above action, the new momentum is 
\begin{eqnarray}
 p_{x'}=x'=(\frac{dt}{d\tau})\dot{x}=[\frac{(\partial_{x}U)^{2}}{4U}]\dot{x}=\frac{\partial_{x}U}{2U^{\frac{1}{2}}}p_{x},
\end{eqnarray}
and we get the trivial symplectic structure
\begin{eqnarray}
 \{x,p_{x'}\}^{*}=\{x,\frac{\partial_{x}U}{2U^{\frac{1}{2}}}p_{x}\}^{*}=1,
\end{eqnarray}
The new Hamiltonian with this reparametrization is
\begin{eqnarray}\label{pseudo}
 H_{x'}=\frac{1}{2}p_{x'}^{2}+\frac{\omega^{2}}{8}(\partial_{x}U)^{2}.
\end{eqnarray}
So, it is possible to establish a relationship, by means of
a gauge transformation, between the harmonic oscillator and a system with arbitrary potential given by $V(x)=\frac{\omega^2}{8}(\partial_{x}U)^{2}$ in one dimension.

 If we consider that $U(x)=2\sqrt{8}x^{\frac{1}{2}}$ where it is noted that
the domain of the function is $(0,\infty)$.
Then the reduced Hamiltonian for this $U(x)$ is
\begin{eqnarray}
 H_{C}=\frac{1}{2}p_{x'}^{2}+\frac{\omega^{2}}{x}.
\end{eqnarray}
Here $\omega$  is an electrical charge,
in particular we are thinking in an electron-electron system.

On the other hand, we can also find a mapping from the complex Harmonic Oscillator
to a Lagrangian that is invariant under the conformal group \cite{fubini}
and resulting a gauge transformation between this conformal Lagrangian
and the real harmonic oscillator.
If we choose $U=\log\mid x\mid$ and using (\ref{lagrangered}) we have
\begin{eqnarray}\label{confa}
 S=\int\!d\tau [\frac{1}{2}(x')^{2}
-\frac{\omega^{2}}{8}\frac{1}{x^{2}}].
\end{eqnarray}
that is conformal invariant.
The Hamiltonian corresponding  using (\ref{pseudo}) is
\begin{eqnarray}
 H_{x'}=\frac{1}{2}p_{x'}^{2}+\frac{\omega^{2}}{8}\frac{1}{x^{2}}.
\end{eqnarray}
In this way we establish a transformation between the real Harmonic
oscillator model and the conformal action (\ref{confa}). 
Furthermore, we may also think in a central potential with a negative charge
so it is necessary to establish another gauge condition
\begin{eqnarray}
\gamma_{-1}=x+iy-iU^{\frac{1}{2}}(y)\approx 0,\label{connorma.11}\\
\{\Phi,\gamma_{-1}\}=-\frac{1}{2U^{\frac{1}{2}}}\partial_{y}U(y),
\end{eqnarray}
In this case the Dirac's bracket is
\begin{eqnarray}
 \{y,p_{y}\}^{*}=\frac{2U^{\frac{1}{2}}(y)}{\partial_{y}U(y)},
\end{eqnarray}
and the resulting momentum generated by this condition is
\begin{eqnarray}
p_{y}=-\frac{\partial_{y}U(y)}{2U^{\frac{1}{2}}(y)}\dot{y}.
\end{eqnarray}
The action associated to the gauge condition (\ref{connorma.11}) will be
\begin{eqnarray}\label{potV2}
 S_{2R}=\int\!dt(\frac{\partial_{y}U}{2U^{\frac{1}{2}}})^{2}[-\frac{1}{2}\dot{y}^{2}-\omega^{2}
\frac{2U^{2}}{(\partial_{y}U)^{2}}],
\end{eqnarray}
and as previously was done, we need a new parameterization
\begin{eqnarray}
 (\frac{d\tau}{dt})=-\frac{4U}{(\partial_{y}U)^{2}}
\end{eqnarray}
resulting a new action from (\ref{potV2}). Now using  $U(y)=2\sqrt{8}y^{\frac{1}{2}}$
we obtain
\begin{eqnarray}
 S=\int\!d\tau [\frac{1}{2}(y')^{2}
+\frac{\omega^{2}}{y}].
\end{eqnarray}
So, the new momentum will be
\begin{eqnarray}
 p_{y'}=y'=(\frac{dt}{d\tau})\dot{y}=-[\frac{(\partial_{y}U(y))^{2}}{4U(y)}]\dot{y}
=\frac{\partial_{y}U}{2U^{\frac{1}{2}}}p_{y},
\end{eqnarray}
and finally we will get the Hamiltonian with opposite sign
\begin{eqnarray}
 H_{C}=\frac{1}{2}p_{y'}^{2}-\frac{\omega^{2}}{y}
\end{eqnarray}
wherewith,  we establish the mapping for this
central potential with opposite sign. In the next section we will show that our idea can be extended to a systems  with more degrees of freedom and that  our transformation is related to  a non-canonical mapping
in two dimensions.

\section{Two dimensional case}
We now describe our strategy for a system with more degrees of freedom.
Specifically, we consider a two dimensional case and we find a gauge transformation
between the harmonic oscillator and a central field.

The initial complex model is given by the Lagrangian
\begin{eqnarray}\label{Lagrangianacom}
L(z_1,z_2)=\frac{1}{2}\dot{z_1}^{2}+\frac{1}{2}\dot{z_2}^{2}-\frac{\omega_{1}^{2}}{2}z_1^{2}-\frac{\omega_{2}^{2}}{2}z_2^{2}
\end{eqnarray}
with $z_1=x+iy$ and $z_2=u+iv$. From, the above Lagrangian we get the momenta
\begin{eqnarray}
p_{x}=\dot{x}+i\dot{y},\qquad
p_{y}=-\dot{y}+i\dot{x},\\
p_{u}=\dot{u}+i\dot{v},\qquad
p_{v}=-\dot{v}+i\dot{u},\nonumber
\end{eqnarray}
resulting into two first class constraints 
\begin{eqnarray}\label{constraingamma}
 \Phi_{0}=p_{x}+ip_{y}\approx 0,\\
 \Phi_{1}=p_{u}+ip_{v}\approx 0.\nonumber
\end{eqnarray} In consequence it is necessary to include two
gauge conditions
\begin{eqnarray}
\gamma_{0}=x+iy-iU^{\frac{1}{2}}(y,v)\approx 0,\label{connorma.111}\\
\gamma_{1}=u+iv-i\mathcal{U}^{\frac{1}{2}}(y,v)\approx 0.\nonumber
\end{eqnarray}
The full set of second class constraints will be $\chi_A=(\Phi_0, \gamma_0, \Phi_1, \gamma_1)$, with the matrix of second class constraints $\mathcal{C}_{AB}=\{ \chi_A, \chi_B\}$ given by

\begin{equation}
\mathcal{C}_{AB}=\left(
\begin{array}{rccccl}
0 & -(\partial_{y}U^{\frac{1}{2}}) & 0 & -(\partial_{y}\mathcal{U}^{\frac{1}{2}})\\
(\partial_{y}U^{\frac{1}{2}})& 0 & (\partial_{v}U^{\frac{1}{2}}) & 0\\
0 & -(\partial_{v}U^{\frac{1}{2}})  & 0 & -(\partial_{v}\mathcal{U}^{\frac{1}{2}})\\
(\partial_{y}\mathcal{U}^{\frac{1}{2}}) & 0 &
(\partial_{v}\mathcal{U}^{\frac{1}{2}}) & 0
\end{array}
\right).
\end{equation}
and the corresponding determinant
\begin{eqnarray}
\det{\mathcal{C}_{AB}}=[(\partial_{y}U^{\frac{1}{2}})(\partial_{v}\mathcal{U}^{\frac{1}{2}})
-(\partial_{y}\mathcal{U}^{\frac{1}{2}})(\partial_{v}U^{\frac{1}{2}})]^{2}\neq 0.
\end{eqnarray}

The total Hamiltonian for this case results
\begin{eqnarray}\label{oaham}
 H_{T(y,v)}=-\frac{1}{2}p_{y}^{2}
+\frac{\omega_{1}^{2}}{2}x^{2}-\frac{\omega_{1}^{2}}{2}y^{2}+i\omega_{1}^{2}xy\\
-\frac{1}{2}p_{v}^{2}
+\frac{\omega_{2}^{2}}{2}u^{2}-\frac{\omega_{2}^{2}}{2}v^{2}+i\omega_{2}^{2}uv\nonumber\\
+\mu^{0} \Phi_{0}+\mu^{1} \Phi_{1}.\nonumber
\end{eqnarray}
In the reduced space the momenta are determined by
\begin{eqnarray}
p_{y}=-[-\dot{y}+i\dot{x}]\mid_{cons}=-\frac{d}{dt}(U^{\frac{1}{2}}),\qquad p_{v}=-[-\dot{v}+i\dot{u}]\mid_{cons}
=-\frac{d}{dt}(\mathcal{U}^{\frac{1}{2}}),
\end{eqnarray}
In the same way we can apply in (\ref{oaham}), the constraints  (\ref{constraingamma})  and the gauge conditions (\ref{connorma.111}) resulting
\begin{eqnarray}
 H_{R}=-\frac{1}{2}p_{y}^{2}-\frac{1}{2}p_{v}^{2}
+\frac{\omega_{1}^{2}}{2}U(y,v)+\frac{\omega_{2}^{2}}{2}\mathcal{U}(y,v),
\end{eqnarray}
By selecting the variables $r^2=y^{2}+v^{2}$ and $y= r \cos(\theta)$ and $v=r \sin(\theta)$ the reduced Lagrangian is
\begin{eqnarray}
L_{R}=-\frac{1}{2}\left[\left(\frac{dU^{\frac{1}{2}}}{dt}\right)^2+ \left(\frac{d\mathcal{U}^{\frac{1}{2}}}{dt}\right)^2 \right]
-\frac{\omega_{1}^{2}}{2}U(r,\theta)-\frac{\omega_{2}^{2}}{2}\mathcal{U}(r,\theta)
\end{eqnarray}
Now by choosing equal frequencies $ \omega_{1}^{2}=\omega_{2}^{2}$ and with the potential given by
\begin{eqnarray}
U^{\frac{1}{2}}=\frac{y}{r^{\beta}}=r^{1-\beta}\cos(\theta),\qquad
\mathcal{U}^{\frac{1}{2}}=\frac{v}{r^{\beta}}=r^{1-\beta}\sin(\theta).
\end{eqnarray}
We obtain for the determinant of the constraint algebra the following expression 
\begin{equation}
\det\mid\mathcal{C}_{AB}\mid=\frac{(2\beta y^{2}+2\beta v^{2}-1)^{2}}{(y^{2}+v^{2})^{4\beta}},
\end{equation}
and the reduced phase-space action is 
\begin{eqnarray}
 S_{R}=\int dt [-\frac{1}{2}r^{2-2\beta}\dot{\theta}^{2}-\frac{(1-\beta)^{2}}{2}r^{-2\beta}\dot{r}^{2}
-\frac{\omega_{1}^{2}}{2}r^{2(1-\beta)}].
\end{eqnarray}
Using a new selection of time
\begin{eqnarray}
(\frac{d\tau}{dt})=r^{2\beta}, \qquad \mathring{A}=\frac{dA}{d\tau},
\end{eqnarray}
and the following change of variables 
\begin{eqnarray}
 \widetilde{\theta}=\frac{1}{(1-\beta)}\theta,\qquad \widetilde{r}=(1-\beta)r,\qquad
\omega_{1}^{2}=\widetilde{\omega}_{1}^{2}(1-\beta)^{2(1-2\beta)}
\end{eqnarray}
we will get the action
\begin{eqnarray}
 \widetilde{S}_{\widehat{R}}=\int d\tau [-\frac{1}{2}\widetilde{r}^{2}\mathring{\widetilde{\theta}}^{2}
-\frac{1}{2}\mathring{\widetilde{r}}^{2}
-\frac{\widetilde{\omega}_{1}^{2}}{2}\widetilde{r}^{2(1-2\beta)}]
\end{eqnarray}
with Hamiltonian

\begin{eqnarray}
 \widetilde{H}_{\widehat{R}}=\frac{1}{2}(p_{\widetilde{r}}^{2}+\frac{p_{\widetilde{\theta}}^{2}}{r^{2}})
-\frac{\widetilde{\omega}_{1}^{2}}{2}\widetilde{r}^{2(1-2\beta)},
\end{eqnarray}
we observe that for $ \beta=\frac{3}{4},$ this Hamiltonian is reduced to the Kepler central problem. So in this way we have shown that our formalism also works for systems in more dimensions. In  next
section we establish how to quantize these systems using path integrals.

\section{Path Integral for the Complex Harmonic Oscillator}
In order to establish the respective quantum mechanics of our complex theories using path integrals  we employ the Senjanovic's method, that allows to quantize systems with second class constraints \cite{Senjanovic}.  
 As an example, we consider the complex harmonic oscillator (\ref{CHO.1}), with gauge condition  $\gamma_{1}$ in the path integral,
and we introduce the notation
\begin{eqnarray}
 \xi^{a}=(x, y),\qquad \rho_{a}=(p_{x}, p_{y}),\\
\tau_{a}=(\Phi,\gamma_{1})
\end{eqnarray}
where $\xi^{a}$ includes the real and imaginary parts of  $z$, and $\rho_{a}$  is in principle a complex quantity.
After that we obtain the measure  of  the path integral that
includes first class constraint and gauge condition and it is
\begin{equation}
 \mathcal{D}\Xi=\mathcal{D}\xi^{a}\mathcal{D}\rho_{a}\det\mid\left\lbrace\tau_{A},\tau_{B}\right\rbrace\mid
\delta(\tau_{C})\\
\end{equation}
and the total path integral is
\begin{eqnarray}
Z=\int\mathcal{D}\Xi \exp [i\int\!dt
(\dot{x}p_{x}+\dot{y}p_{y}-H_{T})].
\end{eqnarray}
If we eliminate $(y,p_{y})$, by means of the delta functionals $\delta(\tau_{C})$ we will get
\begin{eqnarray}
Z_{R}=\int\mathcal{D} x\mathcal{D} p_{x} \prod_{i}|\frac{2U^{\frac{1}{2}}(x_{i})}{\partial_{x_{i}}V(x_{i})}|
|-\frac{i}{2U^{\frac{1}{2}}}\partial_{x_{i}}U(x_{i})|\nonumber\\
\exp [i\int\!dt
(\frac{\partial_{x}U(x)}{2U^{\frac{1}{2}}(x)}\dot{x}p_{x}-H_{R})].
\end{eqnarray}
or using the reparametrization (\ref{repara.1}) we obtain
\begin{eqnarray}
Z_{C}=\int\mathcal{D} x\mathcal{D} p_{x'}
\exp [i\int\!d\tau (\dot{x}p_{x'}-H_{x'})].
\end{eqnarray}
with the Hamiltonian $H_{x'}$ given by (\ref{pseudo}).
In this way, by integrate variables $y, p_{y}$ using  the constraint and gauge condition, we obtain a real path integral.
It is necessary to mention that the complex number don't have ordering, but
we use temporal partitions as a way to order.

\section{Conclusions}\label{conclu}
In this letter, we have shown that there exists in any complex holomorphic system a hidden gauge symmetry. 
This symmetry allows us to map the complex system to several real systems, depending on the gauge condition used.  
By means of the Dirac's method and the Cauchy-Riemann structure, we handle the complex theories establishing  a relationship
with real systems using the gauge symmetry.  In this context, the selection of Hermitian variables,
reality conditions \cite{ashtekar}, or $\mathcal{PT}$ symmetry \cite{Most}, can be
seen as gauge conditions for the first class constraints. 

The procedure established can be summarized as follows:
First,  we start with a complex holomorphic model. As a second step, it is separated in real and imaginary
parts. As a third step, we found primary constraints for the complex momenta and they evolve
correctly in the complex theory, because of the Cauchy-Riemann equations.
As a fourth step, the Hamiltonian is obtained. Finally, we propose gauge conditions as Hermitian conditions and
check the degrees of freedom. Then we have a set of second class constraints and we can quantize this system directly using the path integral. 
For different gauge choices, we can get several real systems that are not related by a canonical transformation in the real phase space but these systems are related by a gauge transformation in the complex phase space. This situation is in some sense analogous to the two time physics of Bars \cite{Bars}, where in the extended phase space physical systems as the free particle and the harmonic oscillator are gauge related. The analysis can also be extended to arbitrary complex dimensions and for systems with high order derivative theories as the Pais-Uhlenbeck model \cite{Margalli.2}. 

\section{Acknowledgments}
The authors acknowledge partial support from CONACyT project 237503 and DGAPA-UNAM grant IN109013.

\end{document}